\documentclass[twocolumn,superscriptaddress,longbibliography]{revtex4-1}
\usepackage{amsmath}
\usepackage{bm}
\usepackage{graphicx}
\usepackage{color} 
\usepackage{setspace}

\newcommand{\ket}[1]{\vert#1\rangle}

\newcommand{\eq}{\begin{equation}}
\newcommand{\fine}{\end{equation}}

\begin{document}

\title{Quantum state fusion in photons}

\author{Chiara Vitelli}
\affiliation{Dipartimento di Fisica, Sapienza Universit\`{a} di Roma, Piazzale Aldo Moro 5, I-00185 Roma, Italy}
\affiliation{Center of Life NanoScience @ La Sapienza, Istituto Italiano di Tecnologia, Viale Regina Elena, 255, I-00185 Roma, Italy}
\author{Nicol\`{o} Spagnolo}
\affiliation{Dipartimento di Fisica, Sapienza Universit\`{a} di Roma, Piazzale Aldo Moro 5, I-00185 Roma, Italy}
\author{Lorenzo Aparo}
\affiliation{Dipartimento di Fisica, Sapienza Universit\`{a} di Roma, Piazzale Aldo Moro 5, I-00185 Roma, Italy}
\author{Fabio Sciarrino}
\affiliation{Dipartimento di Fisica, Sapienza Universit\`{a} di Roma, Piazzale Aldo Moro 5, I-00185 Roma, Italy}
\author{Enrico Santamato}
\affiliation{Dipartimento di Scienze Fisiche, Universit\`{a} di Napoli ``Federico II'',
Complesso Universitario di Monte S. Angelo, via Cintia, 80126 Napoli, Italy}
\author{Lorenzo Marrucci}
\affiliation{Dipartimento di Scienze Fisiche, Universit\`{a} di Napoli ``Federico II'',
Complesso Universitario di Monte S. Angelo, via Cintia, 80126 Napoli, Italy}
\affiliation{CNR-SPIN, Complesso Universitario di Monte S. Angelo, via Cintia, 80126 Napoli, Italy}

\begin{abstract}
Photons are the ideal carriers of quantum information for communication. Each photon can have a single qubit or even multiple qubits encoded in its internal quantum state, as defined by optical degrees of freedom such as polarization, wavelength, transverse modes, etc. Here, we propose and experimentally demonstrate a physical process, named ``quantum state fusion'', in which the two-dimensional quantum states (qubits) of two input photons are combined into a single output photon, within a four-dimensional quantum space. The inverse process is also proposed, in which the four-dimensional quantum state of a single photon is split into two photons, each carrying a qubit. Both processes can be iterated, and hence may be used to bridge multi-particle protocols of quantum information with the multi-degree-of-freedom ones, with possible applications in quantum communication networks.
\end{abstract}
\maketitle

The emerging field of quantum information technology is based on our ability to manipulate and transmit the internal quantum states of physical systems, such as photons, ions, atoms, superconducting circuits, etc.\ \cite{bennett00,ladd10}. In photonic approaches \cite{obrien09,pan12}, much research effort has been recently devoted to expanding the useful quantum space by two alternative approaches: either by increasing the number of involved photons \cite{yao12} or by exploiting different degrees of freedom of the same photon, such as polarization, time-bin, wavelength, propagation paths, and orbital angular momentum \cite{mair01,barreiro05,molina07,lanyon09,ceccarelli09,nagali10pra,straupe10,nagali10prl,gao10,pile11,dada11}. Hitherto, however, these two approaches have proceeded parallel to each other, with little or no attempt at combining them. In this paper, we introduce and prove the realizability of two novel quantum-state manipulation processes, namely fusion and fission, that can be used to bridge these two approaches, allowing for their full integration. Some possible specific applications will be discussed in the concluding paragraphs of this paper.

\begin{figure}
\begin{center}
\includegraphics[width=8.7cm]{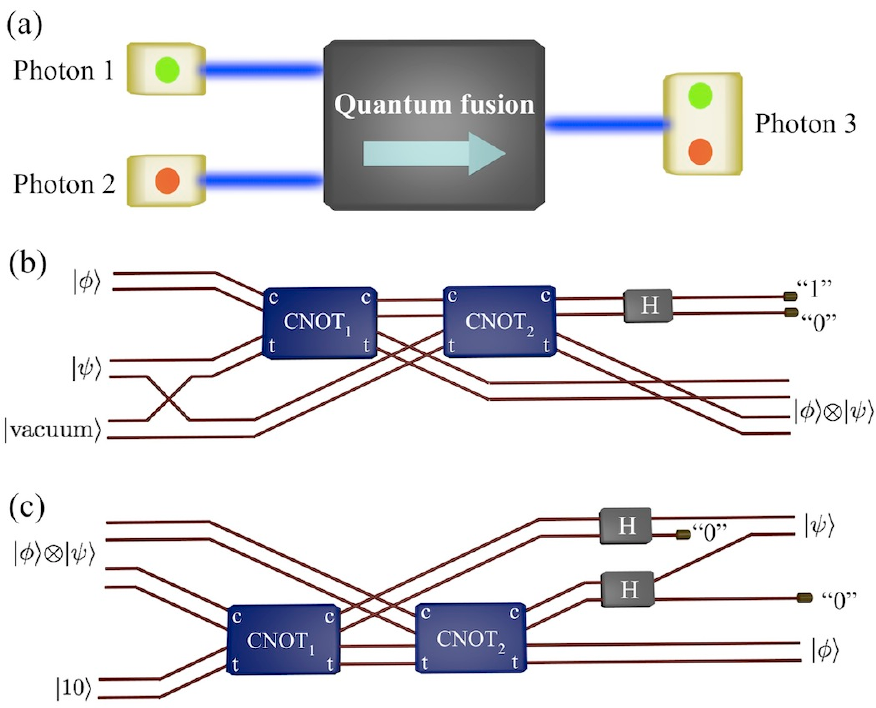}
\end{center}
\caption{\textbf{Quantum state fusion.} (\textbf{a}) Concept: input photons 1 and 2, each carrying a single qubit in their two-dimensional internal states (shown as red and green circles), are merged into output photon 3, carrying both qubits in its four-dimensional internal space. (\textbf{b}) Implementation scheme based on CNOT gates. $\vert\psi\rangle$ and $\vert\phi\rangle$ denote the qubit states of the two input photons. $\vert\psi\rangle\otimes\vert\phi\rangle$ denotes the 2-qubit state of the outgoing photon, conditional on detection of one output photon in the logical-zero output port of the Hadamard gate (H). $c$ and $t$ denote the control and target ports of the CNOT gates. (\textbf{c}) Inverse scheme for quantum state fission. $\vert10\rangle$ denotes the state of an auxiliary input photon that is localized in the upper mode, corresponding to logical-zero. The process success is conditional on detecting no photon in the logical-one output of the two H gates.} \label{figconcept}
\end{figure}

Quantum state fusion is here defined as the physical process by which the internal quantum state of two particles (e.g., photons) is transferred into the four-dimensional internal quantum state of a single particle (photon). In the language of quantum information, two arbitrary input qubits initially attached to separate particles are transferred into a single particle, where they are encoded exploiting at least four independent states of the same particle (i.e., a single-particle ``qudit''). In this framework, this process may also be called ``quantum information fusion'' or ``qubit fusion''. For example, let us assume that initially we have two photons, labeled 1 and 2, having independent polarization quantum states (i.e., carrying polarization-encoded qubits), as follows:
\begin{eqnarray}
\ket{\psi}_1 & = & \alpha \ket{H}_1+\beta \ket{V}_1 \nonumber \\
\ket{\phi}_2 & = & \gamma\ket{H}_2+\delta\ket{V}_2 \label{inputqubits}
\end{eqnarray}
where $H$ and $V$ denote the horizontal and vertical linear polarizations, respectively.
The initial two-photon state can then be written as the product
\begin{eqnarray}
\ket{\Psi}_{12}&=&\left(\alpha\ket{H}_1+\beta\ket{V}_1\right)\otimes
\left(\gamma\ket{H}_2+\delta\ket{V}_2\right) \nonumber\\
&=& \alpha\gamma\ket{H}_1\ket{H}_2+\alpha\delta\ket{H}_1\ket{V}_2+
\beta\gamma\ket{V}_1\ket{H}_2 \nonumber\\
&&+\beta\delta\ket{V}_1\ket{V}_2 \nonumber
\end{eqnarray}
The quantum state fusion corresponds to transforming this state into the same linear combination of four orthogonal single particle states of the outgoing photon 3:
\begin{equation}
\ket{\Psi}_{12} \rightarrow \ket{\Psi}_3 = \alpha\gamma\ket{0}_3+\alpha\delta\ket{1}_3+
\beta\gamma\ket{2}_3+\beta\delta\ket{3}_3 \label{outputqubit}
\end{equation}
where $\ket{0},\ket{1},\ket{2},\ket{3}$ denote four orthogonal internal states of the photon. These four states can be used to define a four-dimensional logical basis of a ``qudit'' carried by the photon. Of course we cannot use the sole two-dimensional polarization space for the outgoing photon. One possibility is to use four independent spatial modes or, alternatively, two spatial modes combined with the two polarizations. Figure \ref{figconcept}a graphically illustrates the quantum fusion concept.

More generally, the fusion process should work even for entangled quantum states, both internally entangled (i.e., the two particles are entangled with each other) and externally entangled (the two particles are entangled with other particles). In the first case, the four coefficients obtained in the tensor product $\alpha\gamma, \alpha\delta, \beta\gamma, \beta\delta$ are replaced with four arbitrary coefficients $\alpha_0, \alpha_1, \alpha_2, \alpha_3$. In the second case, the four coefficients are replaced with four kets representing different quantum states of the external entangled system.

It should be noted that in terms of information content, the quantum fusion has no effect: we have two qubits initially and the same two qubits at the end. In other words, state fusion is not a quantum logical gate for information processing (and it should not be confused with the fusion gate used to create quantum clusters, see e.g. \cite{kok07}). However, physically there is an important transformation, as the two qubits are moved from two separate particles into a single one. This is analogous to what occurs in quantum teleportation, in which the information content is unchanged, but there is a physical transformation in the information localization \cite{bennett93,bouwmeester97,boschi98}.

To be useful in the quantum information field, the process of qubit fusion ideally should be also ``iterable'', i.e. it should be possible to keep adding qubits to the same particle by using a larger and larger internal space. Moreover, the process should be ``invertible'', i.e. one should be able to split a higher-dimensional quantum state of a particle into two (or more) particles. We will name this inverse process ``quantum state fission''.

The feasibility of quantum state fusion and fission in photons is a question of clear fundamental interest, because photons do not interact, or interact very weakly in nonlinear media. Therefore, there is no straightforward way to transfer the quantum state from one photon to the other. The solution we adopt here is based on the Knill-Laflamme-Milburn (KLM) approach to quantum computation with linear optics \cite{knill01,kok07}. A qubit transfer from a particle carrier to another can be generally realized by the application of a single controlled-NOT (CNOT) logical gate. The source qubit in state $\alpha\ket{0}_c+\beta\ket{1}_c$ is used as control (hence the label $c$) and the destination qubit as target, which is initialized to the logical zero $\ket{0}_t$. After the CNOT, the two qubits are entangled in state $\alpha\ket{0}_c\ket{0}_t+\beta\ket{1}_c\ket{1}_t$. A subsequent erasing of qubit $c$ by projection on an unbiased linear combination state $\ket{+}_c=(\ket{0}_c+\ket{1}_c)/\sqrt{2}$ completes the transfer of the quantum state in the qubit $t$. In the case of fusion, however, we have the additional complication that the destination carrier photon also transports another qubit which must not be altered in the CNOT process. Since KLM gates are all based on two-photon interference (i.e. the Hong-Ou-Mandel effect, HOM \cite{hong87}), the presence of the additional qubit makes the two photons partially distinguishable and hence disrupts the gate workings.

\begin{figure}[!]
\begin{center}
\includegraphics[width=0.48\textwidth]{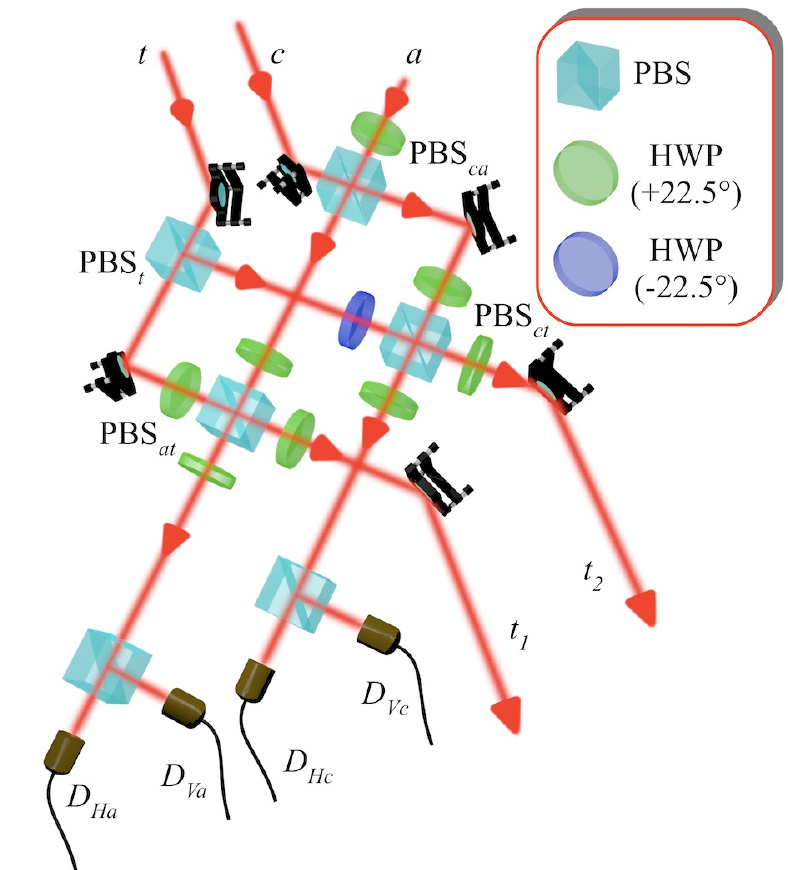}
\end{center}
\caption{\textbf{Schematics of the photon quantum fusion apparatus.} Each red line in this scheme is an optical spatial mode, corresponding to two possible photonic states given by orthogonal polarizations. The input photon quantum states are polarization-encoded (horizontal $H$ and vertical $V$ linear polarizations, corresponding to logical 0 and 1 of photonic qubits, respectively) and travel in spatial modes $c$ (control) and $t$ (target). The ancilla photon must enter the setup in the $H$ polarization state along mode $a$. The output photon ``fused'' quantum state lives in a Hilbert space obtained by combining the two spatial modes $t_1$ and $t_2$ and the two polarizations. The PBSs are assumed to transmit the $H$ polarization and reflect the $V$ one. The intermediate and output modes are given the same label in the transmission through each optical component, except for the unfolding step, realized by PBS$_t$, which splits mode $t$ into modes $t_1$ and $t_2$. HWPs oriented at $22.5^{\circ}$ implement Hadamard gates in the polarization space. The blue-colored HWP is oriented at $-22.5^{\circ}$ so as to be equivalent to a $\hat{\sigma}_{x}$ gate followed by a Hadamard one. PBS$_{ca}$ copies the $c$ photon state onto the ancilla, while PBS$_{at}$ and PBS$_{ct}$ implement the two CNOT gates. Single photon detectors (D) are used to filter the useful output. The fusion is successful if the output $c$ and $a$ photons are both detected in the $H$ channel and there is one and only one photon in each output channel ($c, a, t$), an event which occurs with a probability of 1/32. By exploiting all four possible cases in the $a$ and $c$ polarization detection and a suitable feed-forward, the success probability can be raised to 1/8.} \label{figscheme}
\end{figure}

The solution we found to this problem is based on the trick of initially ``unfolding'' the qubit to be preserved into the superposition of two zero-initialized qubits and then acting on both qubits with a CNOT gate using the same control, as illustrated in Fig.\ \ref{figconcept}b. In this figure, each line represents a possible photonic (spatial and/or polarization) mode, so that a qubit is represented by two lines (this is a ``path'' or ``dual-rail'' encoding of the qubits). We adopt the photon-number notation in order to be able to describe also ``empty'' qubits, i.e., vacuum states, which do occur in the fusion protocol. Given two modes forming a qubit, the $\ket{10}$  ket, where the 0’s and 1’s refer here to the photon numbers, corresponds to having a photon in the first mode, encoding the logical 0 of the qubit. The $\ket{01}$  ket will then represent the photon in the second path, encoding the logical 1 of the qubit. We will also need the $\ket{00}$ ket, representing a vacuum state, i.e. the ``empty'' qubit.
\begin{figure*}[!]
\begin{center}
\includegraphics[width=16cm]{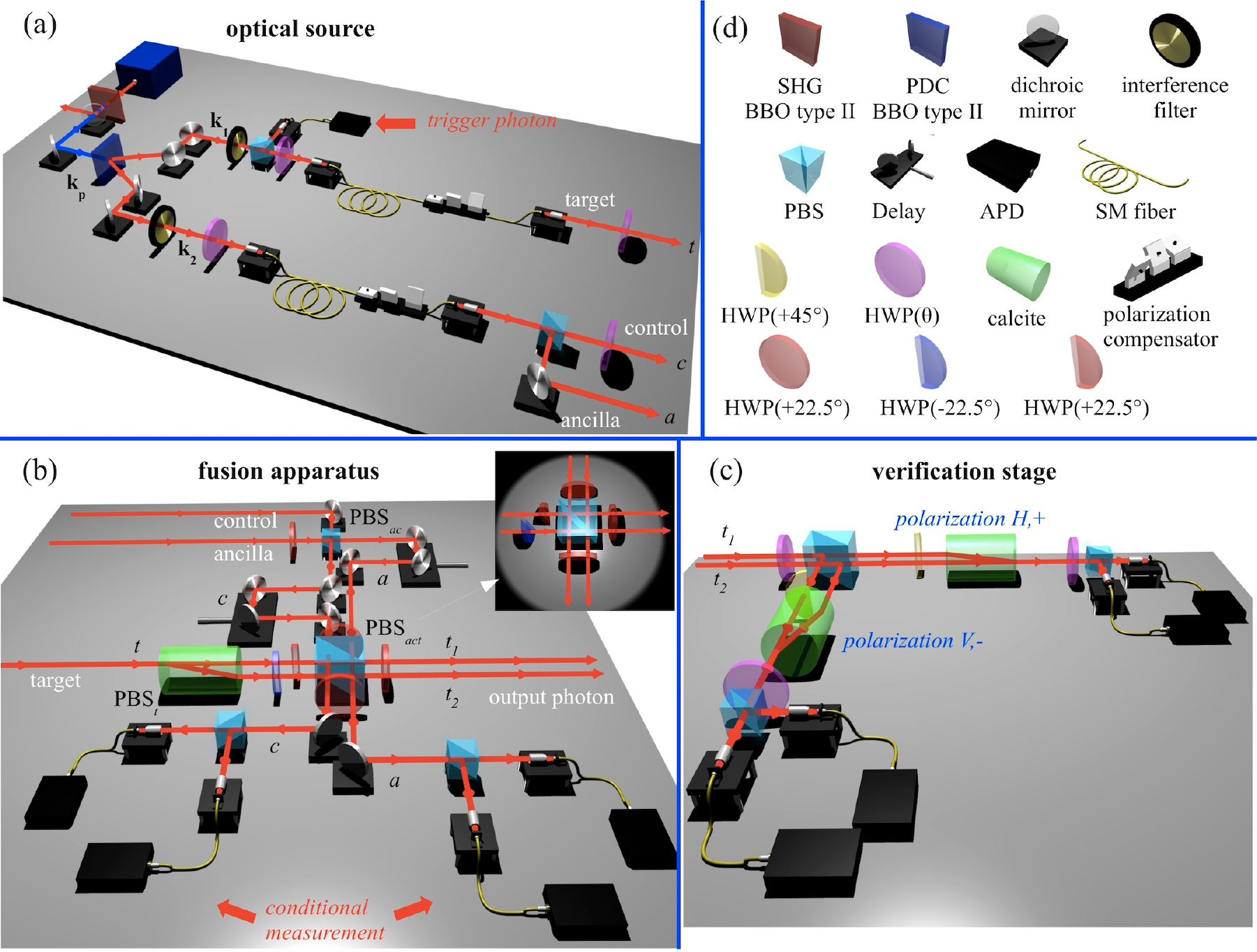}
\end{center}
\caption{\textbf{Experimental setup for implementing and testing the quantum-state fusion.} (\textbf{a}) Optical source used for generating the three input photons. Spontaneous parametric down-conversion (PDC) in a 1.5-mm-thick $\beta$-barium borate crystal (BBO) cut for type-II phase matching, pumped by the second harmonic (SHG) of a Ti:Sa mode-locked laser, generates two polarization-entangled photon pairs at 795 nm (filtered to 3-nm bandwidth) in modes $\mathbf{k}_1$ and $\mathbf{k}_2$ \cite{eibl03}. Each mode is then split by a PBS. One of the four photons is immediately detected by an avalanche photodiode (APD) and used as trigger. The other three are spatially filtered by coupling through single-mode (SM) fibers, have their polarization set by HWPs, and enter the fusion setup along input modes $a, c, t$. (\textbf{b}) Fusion setup, corresponding to the layout shown in Fig.\ \protect\ref{figscheme}. To ensure greater phase stability, the unfolding PBS$_t$ is implemented with a 4-cm-long calcite crystal, so as to obtain two close parallel paths and to use a single PBS$_{act}$ in the place of PBS$_{ct}$ and PBS$_{at}$. Delay lines were used for the time superposition of the photons in the PBSs. (\textbf{c}) Verification stage, used for analyzing the output photon state. The analysis in polarization is made with a HWP and a PBS. Superpositions of modes $t_1$ and $t_2$ were then detected by recombining the two modes in a second calcite crystal and then analyzing again in polarization (this step was not needed for pure $t_1$ or $t_2$ modes). (\textbf{d}) Legend of optical elements. Single-photon count rates were 20-100 kHz, double-concidence rates 200-1000 Hz, and fourfold coincidences few per hour. The degree of photon indistinguishability was quantified via HOM dip visibility \cite{hong87}: we obtained 0.94$\pm$0.01 for photon pairs generated in orthogonal polarization states in modes $\mathbf{k}_1$ and $\mathbf{k}_2$ (belonging to the same PDC pair) and 0.75$\pm$0.05 for photons having the same polarization (different PDC pairs).} \label{figapparatus}
\end{figure*}

The two photons whose state is to be merged travel in two pairs of modes that we will label as $c$ and $t$, with reference to the CNOT input qubits. The initial state of the two photons is taken to be the following (for simplicity we consider the case of two independent input states, but the results are valid also in the more general entangled case):
\begin{eqnarray}
\ket{\psi}_t &=& \alpha\ket{10}_t+\beta\ket{01}_t \nonumber\\
\ket{\phi}_c &=& \gamma\ket{10}_c+\delta\ket{01}_c \label{inputn}
\end{eqnarray}
The qubit ``unfolding'' corresponds to adding two empty modes for photon $t$ and rearranging the four modes so as to obtain the following state:
\begin{equation}
\ket{\psi}_t=\alpha\ket{1000}_t+\beta\ket{0010}_t
\end{equation}
The first two $t$ modes are then treated as one qubit ($t_1$) and the final two $t$ modes as a second qubit ($t_2$), both of them initialized to logical zero, but with the possibility for each of them to be actually empty. Each of these qubits is then subjected to a CNOT with the same $c$ qubit, and finally the $c$ qubit is erased by projection on the $\ket{+}$ state (this corresponds to applying a Hadamard gate and detecting a logical zero). A simple calculation (see Supplementary Materials) shows that this procedure brings the target photon into the desired ``fused'' state
\begin{eqnarray}
\ket{\Psi}_t=(\ket{\psi}\otimes\ket{\phi})_t&=&\alpha\gamma\ket{1000}_t+\alpha\delta\ket{0100}_t \nonumber\\ 
&&+\beta\gamma\ket{0010}_t+\beta\delta\ket{0001}_t. \label{outputn}
\end{eqnarray}
Since the $c$ qubit measurement has a probability of 50\% of obtaining $\ket{+}$, the described method is probabilistic, with a success probability of 50\% (not considering the CNOT success probability). However, the probability can be raised to 100\% by a simple feed-forward procedure, i.e. by applying a suitable unitary transformation on the $t$ photon if the $c$ measurement yields the orthogonal state $\ket{-}=(\ket{0}-\ket{1})/\sqrt{2}$.

One key aspect of the illustrated method is that, after unfolding, each CNOT works with a target photon that carries no additional information, so that interference effects are not disrupted. On the other hand, the CNOT gates must work properly also in the case of empty target qubits, in which case the CNOT must return an unmodified quantum state (see Supplementary Materials for details). This is a non-trivial requirement, not authomatically guaranteed for all KLM approaches, particularly when two CNOTs are used in series, as in the present case.

It can be readily seen that the procedure we have described can be also iterated for merging additional qubits to the same photon. For example, in order to add a third qubit, we only need to insert four additional vacuum modes of the target, organized in four pairs so that all components carrying a nonzero amplitude correspond to the logical 0s of the qubits. Next, we need to perform four CNOT operations always using as control qubit the new qubit to be merged. Finally, we erase the control and leave the eight modes of the target qubit representing the three qubits. And so on. The inverse operation of quantum state fission can be also implemented in a very similar way to the fusion, as shown in Fig.\ \ref{figconcept}c. Further details about this scheme are given in the Supplementary Materials.

Let us now consider a specific demonstrative implementation of the fusion scheme using linear optics, in a KLM approach \cite{knill01}. CNOTs can be implemented in linear optics only probabilistically, although in principle the probability can be made as high as desired by introducing an increasing number of ancilla photons. In our implementation, which uses polarization-encoded qubits, we use the CNOT schemes proposed by Pittman et al., based on half-wave plates (HWP) and polarizing beam-splitters (PBS) \cite{pittman01,pittman03,gasparoni04,zhao05}. Since the $c$ photon must finally be erased, we can use a single photon ancilla for both gates, and the whole scheme becomes actually symmetrical for the exchange of the two CNOTs. The overall fusion scheme, shown in Fig.\ \ref{figscheme}, thus requires three photons only, the two to be merged and the ancilla. Its theoretical success probability is 1/32 without feed-forward and 1/8 with feed-forward and it does not rely on post-selection. The full quantum analysis of this scheme is reported in the Supplementary Materials.

\begin{figure*}
\begin{center}
\includegraphics[width=16cm]{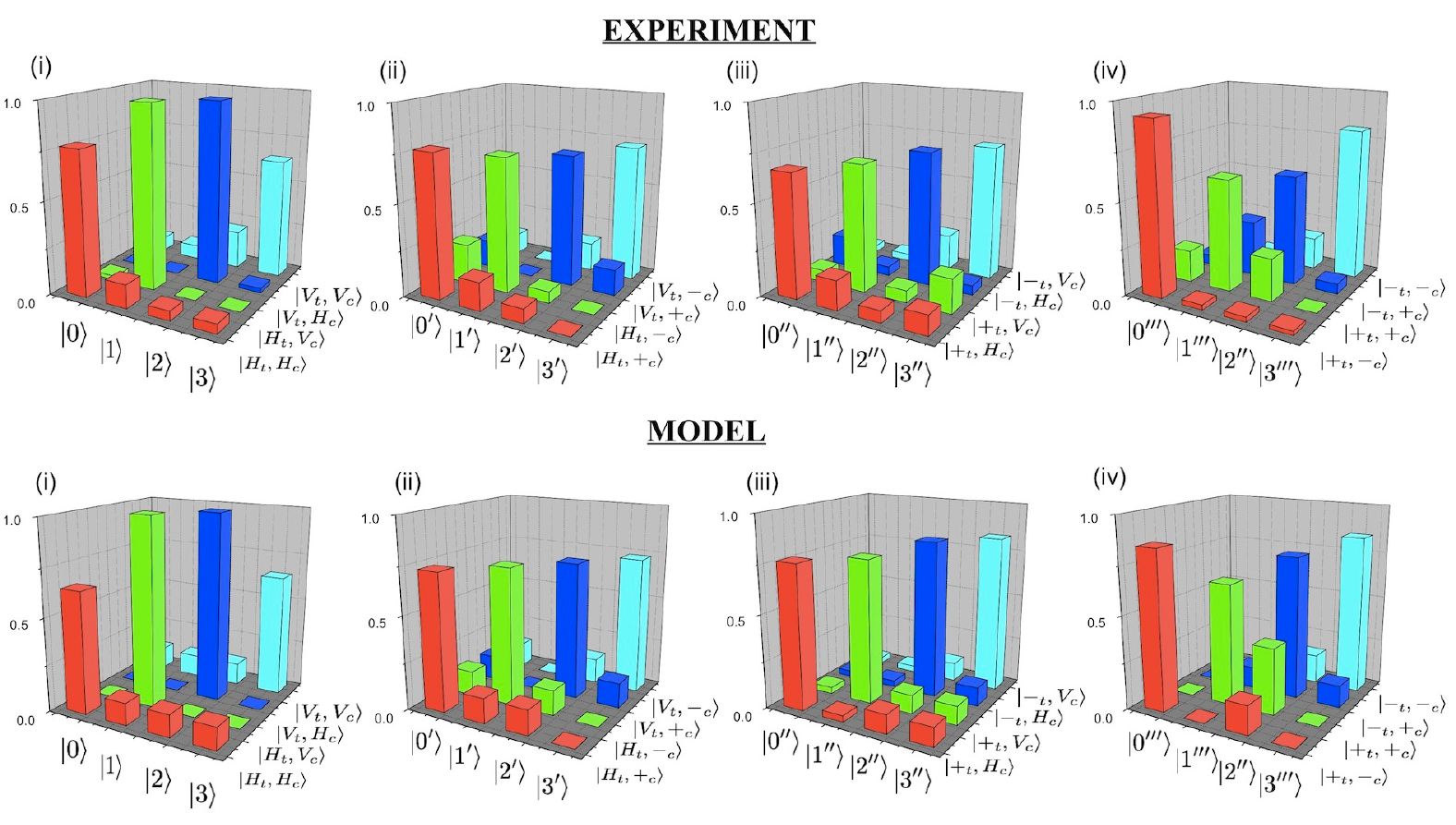}
\end{center}
\caption{\textbf{Experimental results.} The outcome of our quantum fusion experiments (upper panels), compared with the predictions of our model (lower panels) for a degree of photon-pair indistinguishability $p=0.77$. Vertical bars give the fourfold count rates for different input/output combinations. At input, photon-pair states belonging to the following four bases have been used: (i) $\vert H_t, H_c\rangle, \vert H_t, V_c\rangle, \vert V_t, H_c\rangle, \vert V_t, V_c\rangle$, (ii) $\vert H_t, +_c\rangle, \vert H_t, -_c\rangle, \vert V_t, +_c\rangle, \vert V_t, -_c\rangle$, (iii) $\vert +_t, H_c\rangle, \vert+_t, V_c\rangle, \vert-_t, H_c\rangle, \vert-_t, V_c\rangle$, (iv) $\vert+_t, +_c\rangle$, $\vert+_t, -_c\rangle, \vert-_t, +_c\rangle, \vert-_t, -_c\rangle$. At output, the fused-state photon was analyzed in the following corresponding bases (with $n=0,1,2,3$): (i) $\{\vert n\rangle\} = \{\vert H_{t_1}\rangle, \vert V_{t_1}\rangle, \vert H_{t_2}\rangle, \vert V_{t_2}\rangle\}$; (ii) $\{\vert n'\rangle\} = \{\vert +_{t_1}\rangle, \vert -_{t_1}\rangle, \vert +_{t_2}\rangle, \vert -_{t_2}\rangle\}$; (iii) $\{\vert n''\rangle\} = \{\vert H_{t_+}\rangle, \vert V_{t_+}\rangle, \vert H_{t_-}\rangle, \vert V_{t_-}\rangle\}$; (iv) $\{\vert n'''\rangle\} = \{\vert +_{t_+}\rangle, \vert -_{t_+}\rangle, \vert +_{t_-}\rangle, \vert -_{t_-}\rangle\}$. Here, the output modes $t_+$ and $t_-$ are defined according to the rule $\vert H\rangle_{t_\pm}=(\vert H\rangle_{t_1}\pm \vert H\rangle_{t_2})/\sqrt{2}$ and analogously for the other polarizations.} \label{figresultsstates}
\end{figure*}

The experimental apparatus is shown in Fig.\ \ref{figapparatus}. The measurements are based on detecting the fourfold coincidences of the trigger and the output modes $a, c, t$. For testing the fusion apparatus, we performed quantum fusion of $c$ and $t$ photons prepared either in the logical-basis polarization states $\vert H\rangle, \vert V\rangle$ or in their superposition basis states $\vert+\rangle,\vert-\rangle$. The experimental results are shown in the upper panels of Fig.\ \ref{figresultsstates}. For each experimental run, a given state of the two input photons was prepared and the final four-dimensional state of the output photon was measured, by projection on all four basis states belonging to the expected fusion basis (see caption of Fig.\ \ref{figresultsstates}). Ideally, we should find nonzero coincidences only when the detected state is the expected ``fused'' one, corresponding to a unitary fidelity. The measured experimental fidelity, averaged over all tested states, was instead $\mathcal{F}=(75.0\pm1.3$)\%. The average measured fidelities for each tested basis are shown in the Supplementary Materials. All experimental fidelities, both averaged and individual-state ones, are found to be well above the 40\% state-estimation maximal fidelity for a four-dimensional quantum state, showing that our fusion apparatus works much better than a measure-and-prepare trivial approach. Moreover, we could explain quantitatively the non-unitary observed fidelities by considering the fact that the PDC process does not actually generate perfectly identical multiple photon pairs. Therefore, there is always a finite degree of indistinguishability of the three photons that affects the final outcome. We have developed a detailed model of this effect (see Supplementary Materials), whose predictions are shown in the lower panels of Fig.\ \ref{figresultsstates}. The agreement between our model and the experiment is evident. More quantitatively, we evaluated the similarity $\mathcal{S}=\left(\sum_{ij} \sqrt{D_{ij}D^{'}_{ij}}/\sum_{ij}D_{ij}\sum_{ij}D_{ij}^{'}\right)$ between the experimental input/output probability distribution $D_{ij}$ and the predicted one $D'_{ij}$, obtaining $\mathcal{S }=(94.0\pm0.9)\%$. We note that this imperfect indistinguishability of the photon pairs is not a fundamental limitation of the fusion process, and in principle we could improve the fidelity by adopting a narrower spatial and spectral filtering of the PDC output, although at the cost of reducing the coincidence rates.

Before concluding, let us now briefly discuss the application prospects of the quantum-state fusion and fission processes we have introduced here. First, we notice that, although ours has been a bulk-optics demonstration, the fusion/fission processes are expected to find their ideal implementation framework in integrated quantum photonics \cite{sansoni10}. Next, as already mentioned, these processes may enable the interfacing of multi-particle protocols of quantum information, in which different qubits are encoded in different photons, with the multi-degree-of-freedom approaches, in which several qubits are encoded in the same particle. This combination in turn, may have a number of applications, particularly in multi-party quantum communication networks \cite{kimble08}. For example, transmitting a cluster of entangled qubits through a high-loss channel, e.g. an Earth-satellite link, can be done more efficiently by first fusing the qubits in fewer photons, transmitting them, and then splitting them again after detection. Even if the simplest implementations of fusion/fission are not highly efficient, the overall transmission efficiency can still be boosted by a large factor, as it will scale exponentially in the number of transmitted photons. Another example may be the storing of multiple incoming photonic qubits in a smaller number of multi-level matter registers \cite{julsgaard04}. Here a possible advantage would be in the overall rate of decoherence of the stored quantum information, that for entangled states will scale with the number of involved registers. A third example is that of exploiting the multi-degree-of-freedom approach in order to boost the number of qubits that can be processed simultaneously, but without limiting the possibility of interacting with separate parties for the qubit input/output. We also envision many interesting applications in the study of fundamental issues in quantum physics. For example, by exploiting the possibility of fusing states that are entangled with external systems one may create and study complex many-particle clusters of entangled states.

\noindent\textbf{Supplementary Materials} accompanies this manuscript.
\vspace{1 EM}

\noindent\textbf{Acknowledgments}. This work was supported by the Future Emerging Technologies FET-Open Program, within the 7$^{th}$ Framework Programme of the European Commission, under Grant No.\ 255914, PHORBITECH, and by FIRB-Futuro in Ricerca HYTEQ.
\vspace{1 EM}

\noindent\textbf{Author Contributions}. L.M., with contributions from F.S. and E.S., conceived the qubit fusion/fission concept and the corresponding optical schemes. C.V., N.S., and F.S. designed the experimental layout and methodology and, with L.A., carried out the experiments. N.S., C.V., and F.S. developed the model of partial photon distinguishability. All authors discussed the results and participated in drawing up the manuscript.
\vspace{1 EM}

\noindent\textbf{Author Information}. The authors declare no competing financial interests. Correspondence and requests for materials should be addressed to L.M. (lorenzo.marrucci@na.infn.it) or to F.S. (fabio.sciarrino@uniroma1.it).

\clearpage

\onecolumngrid
\appendix
\noindent\textbf{SUPPLEMENTARY MATERIALS}
\vspace{1 EM}

\section{Quantum state fusion}
Let us label as $c$ and $t$ the travelling modes of the two input photons. In a photon-number notation, the two photon states are taken to be the following (for simplicity we consider here the case of two independent input qubits, but the results are valid also in the more general entangled case):
\begin{eqnarray}
\ket{\psi}_t &=& \alpha\ket{10}_t+\beta\ket{01}_t \nonumber\\
\ket{\phi}_c &=& \gamma\ket{10}_c+\delta\ket{01}_c \label{inputn}
\end{eqnarray}
The qubit ``unfolding'' step corresponds to adding two empty modes for photon $t$ and rearranging the four modes so as to obtain the following state:
\begin{equation}
\ket{\psi}_t=\alpha\ket{1000}_t+\beta\ket{0010}_t
\end{equation}
The first two $t$ modes are then treated as one qubit ($t_1$) and the final two $t$ modes as a second qubit ($t_2$), both of them initialized to logical zero, but with the possibility for each of them to be actually empty. Each of these qubits must now be subject to a CNOT gate, using the same $c$ qubit as control. The action of the CNOT gate in the photon-number notation is described by the following equations:
\begin{eqnarray}
\hat{U}_{\text{CNOT}}\ket{10}_c\ket{10}_t&=&\ket{10}_c\ket{10}_t \nonumber\\
\hat{U}_{\text{CNOT}}\ket{01}_c\ket{10}_t&=&\ket{01}_c\ket{01}_t \nonumber\\
\hat{U}_{\text{CNOT}}\ket{10}_c\ket{01}_t&=&\ket{10}_c\ket{01}_t \\
\hat{U}_{\text{CNOT}}\ket{01}_c\ket{01}_t&=&\ket{01}_c\ket{10}_t \nonumber
\end{eqnarray}
However, in the present implementation of the quantum fusion we need to have the CNOT act also on ``empty target qubits'', that is vacuum states. For these we assume the following behavior:
\begin{eqnarray}
\hat{U}_{\text{CNOT}}\ket{10}_c\ket{00}_t&=&\eta\ket{10}_c\ket{00}_t \nonumber\\
\hat{U}_{\text{CNOT}}\ket{01}_c\ket{00}_t&=&\eta\ket{01}_c\ket{00}_t 
\end{eqnarray}
where $\eta$ is a possible complex amplitude rescaling relative to the non-vacuum case. A unitary CNOT must have $|\eta|=1$, but probabilistic implementations do not have this requirement. The qubit fusion scheme works if the two CNOTs have the same $\eta$. In particular the CNOTs implementation proposed by Pittman et al.\ (refs.\ 26-27 of main article) and used in this work have $\eta=1$, so for brevity we will remove $\eta$ in the following expressions.

Let us now consider the entire input state given in Eqs.\ (\ref{inputn}):
\begin{eqnarray}
\ket{\Psi}_i &=& \left(\gamma\ket{10}_c+\delta\ket{01}_c\right)\left(\alpha\ket{1000}_t
+\beta\ket{0010}_t\right) \\
&=& \alpha\gamma\ket{10}_c\ket{10}_{t1}\ket{00}_{t2} +\beta\gamma\ket{10}_c\ket{00}_{t1}\ket{10}_{t2}+\alpha\delta\ket{01}_c\ket{10}_{t1}\ket{00}_{t2} +\beta\delta\ket{01}_c\ket{00}_{t1}\ket{10}_{t2} \nonumber
\end{eqnarray}
where we have also split the overall four-dimensional target state in the two target qubits $t_1$ and $t_2$ which will be subject to the two CNOT gates with the same control $c$. The subsequent application of the two CNOT gates, one acting on $c$ and $t_1$ and the other acting on $c$ and $t_2$, leads to the following state:
\begin{eqnarray}
\ket{\Psi}_f&=&\hat{U}_{\text{CNOT}_2}\hat{U}_{\text{CNOT}_1}\ket{\Psi}_i \nonumber\\
&=& \alpha\gamma\ket{10}_c\ket{10}_{t1}\ket{00}_{t2}+\beta\gamma\ket{10}_c\ket{00}_{t1}\ket{10}_{t2} 
+\alpha\delta\ket{01}_c\ket{01}_{t1}\ket{00}_{t2} +\beta\delta\ket{01}_c\ket{00}_{t1}\ket{01}_{t2} 
\end{eqnarray}
If now we project this state on $\ket{+}_c$, so as to erase the $c$ qubit, and reunite the $t_1$ and $t_2$ kets, we obtain
\begin{equation}
\ket{\Psi}_t=(\ket{\psi}\otimes\ket{\phi})_t = \alpha\gamma\ket{1000}_t+\alpha\delta\ket{0100}_t 
+\beta\gamma\ket{0010}_t+\beta\delta\ket{0001}_t. \label{outputn}
\end{equation}
which is the same as Eq.\ (5) of the main article. Since the $c$ qubit measurement has a probability of 50\% of obtaining $\ket{+}$, without feed-forward the described method has a success probability of 50\% not considering the CNOT success probability.

If the outcome of the $c$ measurement is $\ket{-}_c$, we obtain the following target state:
\begin{equation}
\ket{\Psi}_t=\alpha\gamma\ket{1000}_t-\alpha\delta\ket{0100}_t+
\beta\gamma\ket{0010}_t-\beta\delta\ket{0001}_t.
\end{equation}
This state can be transformed back into Eq.\ (\ref{outputn}) by a suitable unitary transformaton. Therefore, the success probability of the fusion scheme can be raised to 100\% (again not considering CNOTs success probabilities) by a feed-forward mechanism.

\section{Quantum state fusion setup: full calculation}
In this section, the quantum process taking place in the quantum fusion setup is calculated step by step, considering the ideal case of perfectly identical photons. In a following Section, we will consider also the effect of a partial distinguishability of the photons.

The optical layout is shown in figure \ref{figscheme} (or Fig.\ 2 of the main text). We label as $c$ and $t$ the input modes of the two photons to be fused and $a$ the ancilla photon mode. By convention, the intermediate and output modes in the setup are given the same label in the parallel transmission through optical components, except for the unfolding step which splits mode $t$ into modes $t_1$ and $t_2$. The input qubits are polarization-encoded photons, with horizontal $H$ and vertical $V$ linear polarizations standing for logical 0 and 1, respectively. The output ``fused'' photon will be finally encoded in two spatial modes, $t_1$ and $t_2$, and the corresponding $H, V$ polarization modes. The polarizing beam splitters (PBS) are assumed to transmit the $H$ polarization and reflect the $V$ one and not to introduce a phase shift between the two components (a real PBS may possibly introduce a phase shift, but this can be always compensated with suitable birefringent plates, as was actually done in our experiment). The ancilla photon must enter the setup in the $H$ polarization state. The half-wave plates (HWP) are all oriented at 22.5$^\circ$ with respect to the $H$ direction, so as to act as Hadamard gates [$H\rightarrow(H+V)/\sqrt{2}, V\rightarrow(H-V)/\sqrt{2}$], except for the (yellow-colored in the figure) one located after the unfolding PBS along the $t_2$ output mode, which is oriented at  $-22.5^{\circ}$, so as to perform the transformation $V\rightarrow(H+V)/\sqrt{2}$.
%
%
\begin{figure}[!]
\centering
\includegraphics[width=7cm]{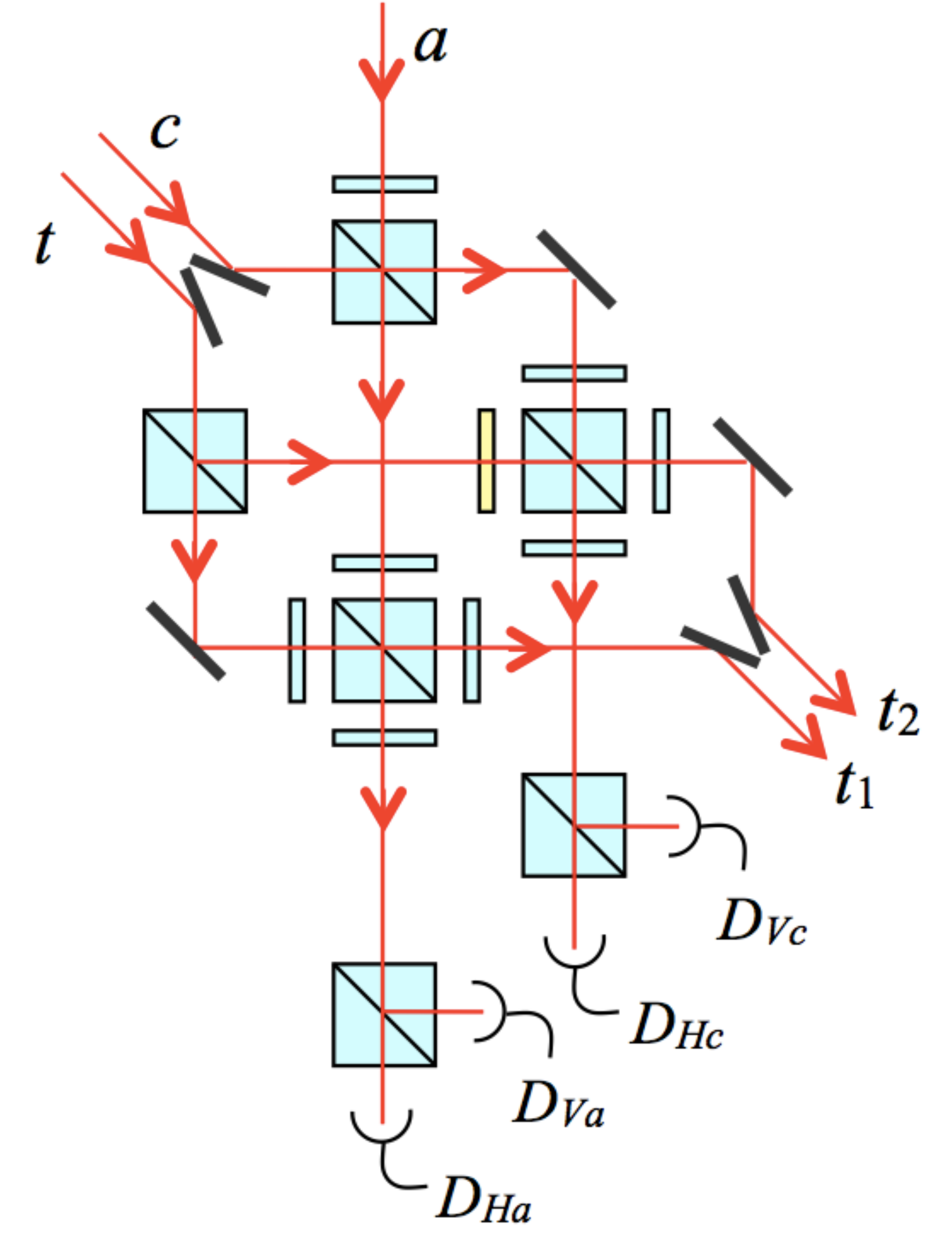}
\caption{Schematics of the quantum fusion apparatus (this figure is equivalent to Fig.\ 2 of the main manuscript and is repeated here for the reader convenience). Each red line in this scheme is a photonic spatial mode, corresponding to two possible photonic states given by orthogonal polarizations. Polarizing beam-splitters (PBS) are indicated by barred squares and assumed to transmit horizontal and reflect vertical polarizations; half-wave plates (HWP) are indicated by thin rectangles and are all oriented at $22.5^{\circ}$ so as to implement Hadamard gates in the polarization space, except for the yellow-colored one which is oriented at $-22.5^{\circ}$ so as to be equivalent to a NOT gate followed by the Hadamard one; single photon detectors (D) are used to filter the useful output.} \label{figscheme}
\end{figure}

Let us now calculate the behavior of our fusion apparatus. The three-photon input state is the following:
\begin{equation}
\vert \Psi_i \rangle = (\alpha_0 H_t H_c + \alpha_1 H_t V_c + \alpha_2 V_t H_c + \alpha_3 V_t V_c)H_a
\end{equation}
where in the right-hand-side we omit the ket symbols for brevity and we have assumed the $c$ and $t$ photons to be in an arbitrary two-photon state, either separable or entangled. In particular, the case of separable qubits considered in the main article corresponds to setting $\alpha_0=\alpha\gamma, \alpha_1=\alpha\delta, \alpha_2=\beta\gamma, \alpha_3 = \beta\delta$, where $(\alpha,\beta)$ and $(\gamma,\delta)$ are the coefficient pairs defining the two qubit states. As we shall see below, the symbols $H$ and $V$ should be actually taken to represent the photon creation operators (for the given polarization and spatial mode) to be applied to the vacuum state, in order to obtain the resulting ket state. The creation-operator interpretation is needed for the states involving two or more photons in a same mode, which in our case appear only as intermediate states but not in the final output.

We now consider the effect of each optical component in sequence, indicating as subscript(s) the mode(s) on which the optical component is acting:
\begin{equation}
\xrightarrow{\text{HWP}_a} (\alpha_0 H_t H_c + \alpha_1 H_t V_c + \alpha_2 V_t H_c + \alpha_3 V_t V_c)\frac{1}{\sqrt{2}}(H_a+V_a) \nonumber
\end{equation}
\begin{equation}
\xrightarrow{\text{PBS}_{a,c}}\frac{1}{\sqrt{2}}[(\alpha_0 H_t + \alpha_2 V_t) H_c H_a + (\alpha_1 H_t + \alpha_3 V_t) V_c V_a + (\alpha_1 H_t +\alpha_3 V_t) H_a V_a + (\alpha_0 H_t + \alpha_2 V_t) H_cV_c)] \nonumber
\end{equation}
\begin{eqnarray}
\xrightarrow{\text{HWP}_a\text{HWP}_c}&&\frac{1}{2\sqrt{2}}[(\alpha_0 H_t + \alpha_2 V_t) (H_c + V_c) (H_a + V_a) + (\alpha_1 H_t + \alpha_3 V_t) (H_c-V_c) (H_a-V_a)\nonumber\\
&& + (\alpha_1 H_t +\alpha_3 V_t) (H_a^2- V_a^2) + (\alpha_0 H_t + \alpha_2 V_t) (H_c^2-V_c^2)] \nonumber
\end{eqnarray}
where in the last expression, we introduced the squared symbols $H^2$ or $V^2$ to denote a two-photon state for the given mode (the precise normalization convention here corresponds to considering the $H$ and $V$ symbols as creation operators for the given mode acting on vacuum, or equivalently to setting $H^2 = \sqrt{2}\vert 2 \rangle_H$ in the photon-number ket notation, and similar). Continuing with the unfolding step in the $t$ mode:
\begin{eqnarray}
\xrightarrow{\text{PBS}_t}&&\frac{1}{2\sqrt{2}}[(\alpha_0 H_{t_1} + \alpha_2 V_{t_2}) (H_c + V_c) (H_a + V_a) + (\alpha_1 H_{t_1} + \alpha_3 V_{t_2}) (H_c-V_c) (H_a-V_a) \nonumber\\
&& + (\alpha_1 H_{t_1} +\alpha_3 V_{t_2}) (H_a^2- V_a^2) + (\alpha_0 H_{t_1} + \alpha_2 V_{t_2}) (H_c^2-V_c^2)] \nonumber
\end{eqnarray}
Note that the absence of a symbol for a given mode in a term corresponds to having the vacuum state in that mode (this again corresponds to the creation-operator notation). Next, we have
\begin{eqnarray}
\xrightarrow{\text{HWP}_{t_1}\text{HWP}_{t_2}}&&\frac{1}{4}[(\alpha_0 H_{t_1}+\alpha_0 V_{t_1}+\alpha_2 H_{t_2}+\alpha_2 V_{t_2}) (H_c + V_c) (H_a + V_a)  \nonumber\\
&& + (\alpha_1 H_{t_1}+\alpha_1 V_{t_1}+\alpha_3 H_{t_2} +\alpha_3 V_{t_2}) (H_c-V_c) (H_a-V_a)\nonumber\\
&& + (\alpha_1 H_{t_1}+\alpha_1 V_{t_1}+\alpha_3 H_{t_2} +\alpha_3 V_{t_2}) (H_a^2- V_a^2)+ (\alpha_0 H_{t_1}+\alpha_0 V_{t_1}+\alpha_2 H_{t_2}+\alpha_2 V_{t_2}) (H_c^2-V_c^2)]  \nonumber
\end{eqnarray}
Let us now consider the effect of each of the two PBS acting as CNOTs. We will also drop all terms which do not lead to one (and only one) output photon in each of the two modes $c$ and $a$. The latter is a projection step (PS) ensured by the final photon detection. We thus obtain
\begin{eqnarray}
\xrightarrow{\text{PBS}_{a,t_1}\text{PS}_a}&&\frac{1}{4}[(\alpha_0 H_{t_1}H_a+\alpha_0 V_{t_1}V_a+\alpha_2 H_{t_2}H_a+\alpha_2 V_{t_2}H_a) (H_c + V_c)\nonumber\\
&& + (\alpha_1 H_{t_1}H_a-\alpha_1 V_{t_1}V_a+\alpha_3 H_{t_2}H_a +\alpha_3 V_{t_2}H_a) (H_c-V_c)+ \alpha_0 V_a (H_c^2-V_c^2)] \nonumber
\end{eqnarray}
\begin{eqnarray}
\xrightarrow{\text{PBS}_{c,t_2}\text{PS}_c}&&\frac{1}{4}(\alpha_0 H_{t_1}H_aH_c+\alpha_0 V_{t_1}V_aH_c+\alpha_2 H_{t_2}H_aH_c +\alpha_2 V_{t_2}H_aV_c+\alpha_1 H_{t_1}H_aH_c-\alpha_1 V_{t_1}V_aH_c  \nonumber\\
&& +\alpha_3 H_{t_2}H_aH_c -\alpha_3 V_{t_2}H_aV_c) \nonumber
\end{eqnarray}
and finally all the exit half-wave plates:
\begin{eqnarray}
\!\!\!\!\xrightarrow{\text{HWP}_{a}\text{HWP}_{c}\text{HWP}_{t_1}\text{HWP}_{t_2}}&&\frac{1}{8\sqrt{2}}[(\alpha_0\!+\!\alpha_1) (H_{t_1}+V_{t_1})(H_a+V_a)(H_c+V_c) +(\alpha_0\!-\!\alpha_1)(H_{t_1}-V_{t_1})(H_a-V_a)(H_c+V_c)  \nonumber\\
&&+(\alpha_2\!+\!\alpha_3)(H_{t_2}+V_{t_2})(H_a+V_a)(H_c+V_c) +(\alpha_2\!-\!\alpha_3) (H_{t_2}-V_{t_2})(H_a+V_a)(H_c-V_c)\nonumber
\end{eqnarray}
By expanding the products, we obtain the following final state:
\begin{eqnarray}
\vert \Psi_f \rangle &=& \frac{H_aH_c}{4\sqrt{2}}(\alpha_0 H_{t_1}+\alpha_1 V_{t_1} + \alpha_2 H_{t_2} + \alpha_3 V_{t_2}) + \frac{H_aV_c}{4\sqrt{2}}(\alpha_0 H_{t_1}+\alpha_1 V_{t_1} + \alpha_2 V_{t_2} + \alpha_3 H_{t_2})\nonumber\\
&+&\frac{V_aH_c}{4\sqrt{2}}(\alpha_0 V_{t_1}+\alpha_1 H_{t_1} + \alpha_2 H_{t_2} + \alpha_3 V_{t_2}) + \frac{V_aV_c}{4\sqrt{2}}(\alpha_0 V_{t_1}+\alpha_1 H_{t_1} + \alpha_2 V_{t_2} + \alpha_3 H_{t_2})\nonumber
\end{eqnarray}
It is clear from this expression that if the detectors of the $a$ mode and $c$ mode both detect $H$ photons, then in the $t$ modes we obtain directly the merged photon (this occurs with a probability of 1/32):
\begin{equation}
\vert \Psi_f \rangle = \alpha_0 H_{t_1}+\alpha_1 V_{t_1} + \alpha_2 H_{t_2} + \alpha_3 V_{t_2}
\end{equation}
The other cases can be transformed into this by a simple feed-forward mechanism: a $V$ detection on the $a$ channel will indicate that we must apply a $\hat{\sigma}_x$ operator on the $t_1$ mode, while a $V$ detection on the $c$ channel indicates that we must apply it on the $t_2$ mode.

By exploiting all cases, we have a total success probability for the fusion of 1/8. In principle, the scheme may become quasi-deterministic by adopting a KLM-like approach with teleportation and a very large number of ancilla photons.

It is remarkable that we could apply the one-ancilla CNOT in the first step and the outcome was not ruined by the presence of intermediate double-photon modes. This occurs because the double-photon modes running in the $a$ or $c$ channels always have two photons with orthogonal polarizations. The half-wave plates will then transform them into $HH$ or $VV$ double photons, which are either both reflected or both transmitted at the following PBS, so that the final measurement-filtering stage will not include these cases. In other words, the polarization Hong-Ou-Mandel effect will make sure that these ``wrong'' channels do not contribute to the final projected state.

Another remarkable fact is that in principle the projection step does not require a destructive post-selection if the detectors are able to distinguish between one- and two-photon events, as in this case the detection of the $t$ photon is not necessary to determine the success of the process.

\section{Quantum state fission}
The general scheme of the quantum state fission process is given in Fig.\ 1c of the main article. Adopting the photon number notation, we assume to have an input photon encoding two qubits in the four-path state
\begin{equation}
\vert \psi_i \rangle = \alpha_0\vert1000\rangle+\alpha_1\vert0100\rangle+\alpha_2\vert0010\rangle +\alpha_3\vert0001\rangle
\end{equation}
We label this input photon as $c$ (for control). We also label the first two modes as $c_1$ and the last two modes as $c_2$. We also have another photon, labeled as $t$ (for target), that is initialized in the logical zero state of two other modes, so that the initial two-photon state is the following:
\begin{equation}
\vert \Psi_i \rangle = (\alpha_0\vert10\rangle_{c_1} \vert00\rangle_{c_2}  +\alpha_1\vert01\rangle_{c_1}\vert00\rangle_{c_2}+\alpha_2\vert00\rangle_{c_1}\vert10\rangle_{c_2} +\alpha_3\vert00\rangle_{c_1}\vert01\rangle_{c_2})\vert10\rangle_t \nonumber
\end{equation}
We now apply the two CNOT gates in sequence, using the $t$ photon as target qubit in both cases and the $c_1$ and $c_2$ modes of the $c$ photon as control qubit in the first and second CNOT, respectively. In order to do these operations properly, we need to define the CNOT operation also for the case when the control qubit is empty. As for the previous case of empty target qubit, the CNOT outcome in this case is taken to be simply identical to the input except for a possible amplitude rescaling, i.e.
\begin{eqnarray}
\hat{U}_{\text{CNOT}}\vert00\rangle_c\vert10\rangle_t &=& \eta\vert00\rangle_c\vert10\rangle_t \nonumber\\
\hat{U}_{\text{CNOT}}\vert00\rangle_c\vert01\rangle_t &=& \eta\vert00\rangle_c\vert01\rangle_t
\end{eqnarray}
which is what occurs indeed in most CNOT implementations. Our fission scheme works well if the two CNOTs have the same $\eta$ factor. For brevity we simply assume $\eta=1$ in the following, which is the case of the CNOT implementations we will utilize. Hence, we obtain
\begin{equation}
\hat{U}_{\text{CNOT}_1}\hat{U}_{\text{CNOT}_2}\vert \Psi_i \rangle = \alpha_0\vert10\rangle_{c_1} \vert00\rangle_{c_2} \vert10\rangle_t + \alpha_1\vert01\rangle_{c_1} \vert00\rangle_{c_2} \vert01\rangle_t+\alpha_2\vert00\rangle_{c_1}\vert10\rangle_{c_2}\vert10\rangle_t +\alpha_3\vert00\rangle_{c_1} \vert01\rangle_{c_2} \vert01\rangle_t\nonumber
\end{equation}
Now we need to erase part of the information contained in the control photon. This is accomplished by projecting onto $\vert + \rangle$ combinations of the first and second pairs of modes, while keeping unaffected their relative amplitudes. In other words, as shown Fig.\ 1c of the main article, we must apply an Hadamard transformation on both pairs of modes, and take as successful outcome only the logical-zero output (corresponding to the $\vert + \rangle$ combination of the inputs). The projection is actually performed by checking that no photon comes out of the $\vert - \rangle$ (i.e., logical one) output ports of the Hadamard. The two surviving output modes are then combined into a single output $c$-photon qubit, which together with the $t$-photon qubit form the desired split-qubit output. Indeed, we obtain the following projected output:
\begin{equation}
\vert\Psi_f\rangle = \alpha_0\vert10\rangle_{c} \vert10\rangle_t + \alpha_1\vert10\rangle_{c}\vert01\rangle_t + \alpha_2\vert01\rangle_{c}\vert10\rangle_t +\alpha_3\vert01\rangle_{c} \vert01\rangle_t
\end{equation}
which describes the same two-qubit state as the input, but encoded in two photons instead of one.

The proposed scheme for fission has a 50\% probability of success, not considering the CNOT contribution. 
It might be again possible to bring the probability to 100\% (not considering CNOTs) by detecting the actual $c$-photon output mode pair after the Hadamard gates by a quantum non-demolition approach or in post-selection, and then applying an appropriate unitary transformation to the $t$ photon. Alternatively, one can in principle use the KLM approach with many ancilla photons and teleportation to turn the scheme into a quasi-deterministic one.

%
%
\begin{figure}[ht!]
\centering
\includegraphics[width=7.5cm]{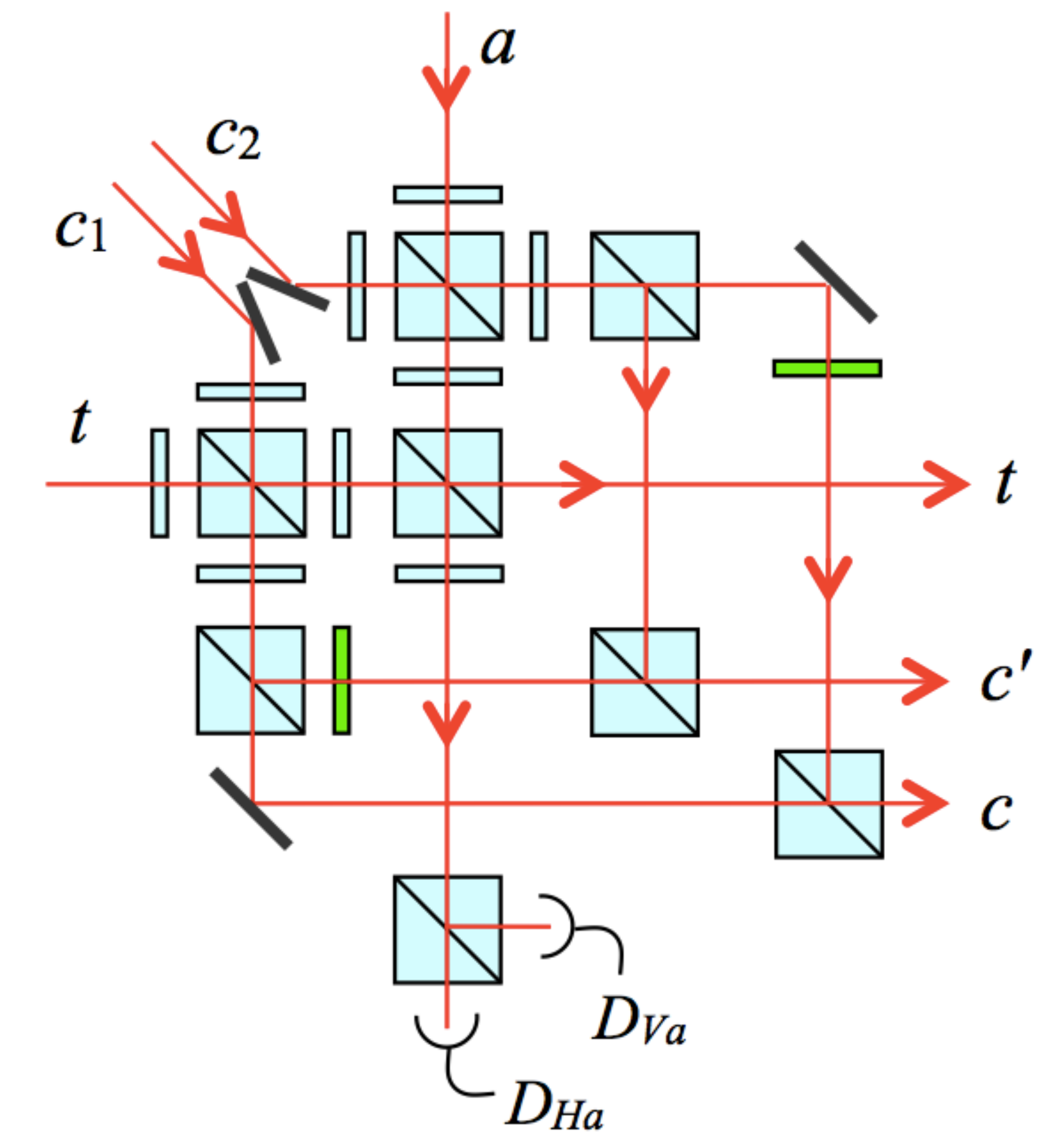}
\caption{Schematics of the quantum fission apparatus. As in Fig.\ S1, each red line in this scheme is a photonic spatial mode, corresponding to two possible photonic states given by orthogonal polarizations. The input photon enters in the modes $c_1$-$c_2$. Splitted ouput photons exit along mode $t$ and either $c$ or $c'$. Barred squares indicate PBSs; thin rectangles indicate HWPs, all oriented at $22.5^{\circ}$ except for the green-colored one which are oriented at $45^{\circ}$ so as to be equivalent to a NOT gate swapping $H$ and $V$.} \label{figsplittingscheme}
\end{figure}
A possible specific apparatus for quantum state fission is shown in Fig.\ S2. In this setup, output photon internal quantum states are defined in the polarization space, as for the fusion setup case. The four modes of input photon $c$ are defined by the two spatial modes $c_1$ and $c_2$ and the two polarizations $H$ and $V$. Both the target photon $t$ and the ancilla photon $a$ must enter the setup after being initialized to logical zero, i.e. $H$-polarized.

The input three-photon state is the following:
\begin{equation}
\vert \Psi_i \rangle = (\alpha_0 H_{c_1} + \alpha_1 V_{c_1} + \alpha_2 H_{c_2} + \alpha_3 V_{c_2}) H_t H_a,
\end{equation}
The calculations of the optical component effect on this state can be carried out similarly to the case of the fusion setup and we will not repeat them here. We give only the final output obtained after all components and after projection on a subspace in which there is only one photon per output mode (i.e., one in $a$, one in $t$, and one in either $c$ or $c'$):
\begin{eqnarray}
\vert \Psi_f \rangle &=& \frac{H_a}{4\sqrt{2}}(\alpha_0 H_tH_c+\alpha_1 V_tH_c + \alpha_2 H_tV_c + \alpha_3 V_tV_c) +\frac{V_a}{4\sqrt{2}}(\alpha_0 H_tH_c-\alpha_1 V_tH_c + \alpha_2 H_tV_c - \alpha_3 V_tV_c)\nonumber\\
&+&\frac{H_a}{4\sqrt{2}}(\alpha_0 V_tH_{c'}+\alpha_1 H_tH_{c'} + \alpha_2 V_tV_{c'} + \alpha_3 H_tV_{c'}) +\frac{V_a}{4\sqrt{2}}(-\alpha_0 V_tH_{c'}+\alpha_1 H_tH_{c'} - \alpha_2 V_tV_{c'} + \alpha_3 H_tV_{c'})
\nonumber
\end{eqnarray}
This expression shows that again the success probability is of 1/32 without feed-forward. If we detect the ancilla photon in the $H$ polarization and a photon coming out on the $c$ channel, then the final state is the desired one:
\begin{equation}
\vert \Psi_f \rangle = \alpha_0 H_tH_c+\alpha_1 V_tH_c + \alpha_2 H_tV_c + \alpha_3 V_tV_c
\end{equation}
Also in this case a feed-forward mechanism can increase the success probability. In particular, if the output ancilla is $V$-polarized, then we must change the sign of the $V_t$ mode. If the control photon comes out from the $c'$ channel, instead of the $c$ channel, then we must swap $V_t$ and $H_t$. The ancilla detection can be done without disturbing the $c$ and $t$ output photons, so with a real feed-forward mechanism we can double the success probability. The control photon detection can only be done destructively (not considering quantum non-demolition possibilities) and hence a true feed-forward is not practical (in principle we could post-pone the measurement of the target photon and correct its polarization state, but in practice this is not useful). Once again, however, a KLM-like approach based on a large number of ancilla photons can in principle turn this process into a quasi-deterministic one.

\section{Modeling photon distinguishability}
The three-photon state used as input in the quantum fusion experiment was conditionally prepared by exploiting the second-order process of a type-II spontaneous PDC source, generating two photon pairs, with one of the four photons used only as trigger. A crucial parameter in the four-photon component of the generated state is given by the partial distinguishability of the two generated photon-pairs. In this section we develop a theoretical model for the experiment which takes into account an imperfect indistinguishability between the three photons. To model the effect of spectral and temporal distinguishability, the generated four-photon density matrix in the two output spatial modes $\mathbf{k}_{1}$ and $\mathbf{k}_{2}$ can be written as follows\cite{Tsuj04}:
\begin{equation}
\varrho= p \vert \psi_{-}^{4} \rangle \langle \psi_{-}^{4} \vert + (1-p) \vert \psi_{-}^{2} \rangle^{(A)} \langle \psi_{-}^{2} \vert \otimes \vert \psi_{-}^{2} \rangle^{(B)} \langle \psi_{-}^{2} \vert,
\end{equation}
where:
\begin{eqnarray}
&& \vert \psi_{-}^{4} \rangle = \frac{1}{\sqrt{3}} (\vert H H \rangle_{1} \vert V V \rangle_{2} - \vert H V \rangle_{1} \vert H V \rangle_{2} + \vert V V \rangle_{1} \vert H H \rangle_{2}),\\
&& \vert \psi_{-}^{2} \rangle^{(i)} = \frac{1}{\sqrt{2}} (\vert H \rangle_1^{(i)} \vert V \rangle_2^{(i)} - \vert V \rangle_1^{(i)}\vert H \rangle_2^{(i)}).
\end{eqnarray}
Here, the parameter $p$ is the fraction of events in which two indistinguishable photon pairs are generated in the $\vert \psi_{-}^{4} \rangle$ state. Conversely, $(1-p)$ is the fraction of events in which two distinguishable photon pairs are emitted by the source (labeled by the superscript $i=A,B$). Hence, the parameter $p$ represents the degree of indistinguishability of the four-photon state. Referring to the experimental scheme of Fig.\ 3 in the main text, the output state is analyzed in polarization by selecting the contributions in which two orthogonally-polarized photons are generated on each spatial mode. The corresponding weights on the states $\vert \psi_{-}^{4} \rangle$ and $\vert \psi_{-}^{2} \rangle^{(A)} \otimes \vert \psi_{-}^{2} \rangle^{(B)}$ are respectively $1/3$ and $1/2$, leading to an overall fraction of events with three indistinguishable photons given by $r=2p/(3-p)$. 

Let us now proceed with the evolution induced by the quantum fusion apparatus. Without loss of generality, we consider the case in which the photons in the ancillary output modes are projected in the $\vert H \rangle$ polarization state. We first observe that due to the polarization analysis of the source, the control and the target qubit always belong to the same photon pair since they are selected with orthogonal polarizations in the two different spatial modes. When two indistinguishable photon-pairs are produced by the source, the input state in the apparatus takes the form
\begin{equation}
\vert \psi_{\mathrm{ind}} \rangle = H_a (\alpha H_t + \beta V_t) (\gamma H_c + \delta V_c) ,
\end{equation}
where we have resumed here the shortened notation in which the ket symbols on the left-hand-side are omitted. In this case, with probability $P_{\mathrm{ind}} = 1/32$, the output photon will emerge only in the desired output state, corresponding to the qubit fusion. When the two pairs emitted by the source are distinguishable, instead, the ancillary photon belongs to a different photon pair with respect to the control and target photons. The input state in the qubit fusion apparatus can then be written as follows:
\begin{equation}
\vert \psi_{\mathrm{d}} \rangle = H_a^{(A)} (\alpha H_t^{(B)} + \beta V_t^{(B)})(\gamma H_c^{(B)} + \delta V_c^{(B)}) 
\end{equation}
where the superscripts $(A)$ and $(B)$ distinguish the photons and hence remove the interference effects between the ancilla and the other two photons, thus leading to spurious contributions in the output state and decreasing the fidelity. The overall input state is then obtained as:
\begin{equation}
\varrho_{p} = r \vert \psi_{\mathrm{ind}} \rangle \langle \psi_{\mathrm{ind}} \vert + (1-r) \vert \psi_{\mathrm{d}} \rangle \langle \psi_{\mathrm{d}} \vert
\end{equation}
The output state can be evaluated by applying to this input the same sequence of operations of the fusion apparatus reported above and finally keeping only the terms corresponding to the detection of two horizontally-polarized photons in the ancilla and control output modes.

We skip the intermediate calculations and in the following just give the results we have obtained for the four state bases which were tested in our quantum fusion experiment.

\textbf{Basis (i)}, using $\{ \vert H \rangle_{t}, \vert V \rangle_{t}; \vert H \rangle_{c},\vert V \rangle_{c}\}$. The input/output probability matrix for this basis reads:
\begin{equation}
P_{p}^{(i)} = \begin{pmatrix}
\frac{3+p}{12-8p} &\frac{3-p}{12-8p} &\frac{3-p}{12-8p} &\frac{3-p}{12-8p} \\
0 &1 &0 &0 \\
0 &0 &1 &0 \\
\frac{3-p}{12-8p} &\frac{3-p}{12-8p} &\frac{3-p}{12-8p} &\frac{3+p}{12-8p}
\end{pmatrix}.
\end{equation}
The matrix rows correspond to the following input two-photon states: $\{ \vert H_t H_c\rangle, \vert H_t V_c \rangle, \vert  V_t H_c \rangle, \vert V_t V_c\rangle\}$. The matrix columns correspond to the following single-photon output states: $\{ \vert 0 \rangle = \vert H\rangle_{t_1}, \vert 1 \rangle = \vert V\rangle_{t_1}, \vert 2 \rangle = \vert H\rangle_{t_2}, \vert 3 \rangle = \vert V\rangle_{t_2} \}$. Of course, the matrix is normalized so that the sum for each row equals unity.

\textbf{Basis (ii)}, using $\{ \vert H \rangle_{t}, \vert V \rangle_{t}; \vert + \rangle_{c},\vert - \rangle_{c} \}$. The input/output probability matrix reads:
\begin{equation}
P_{p}^{(ii)} = \begin{pmatrix}
\frac{3+p}{9-5p} &\frac{3-p}{9-5p} &0 &\frac{3-p}{9-5p} \\
\frac{3-p}{9-5p} &\frac{3+p}{9-5p} &0 &\frac{3-p}{9-5p} \\
0 &\frac{3-p}{9-5p} &\frac{3+p}{9-5p} &\frac{3-p}{9-5p} \\
0 &\frac{3-p}{9-5p} &\frac{3-p}{9-5p} &\frac{3+p}{9-5p}
\end{pmatrix}.
\end{equation}
The rows correspond to the following input two-photon states: $\{ \vert H_t +_c\rangle, \vert H_t -_c\rangle, \vert V_t +_c\rangle, \vert V_t -_c\rangle\}$. The columns correspond to the following single-photon output states: $\{ \vert 0' \rangle = \vert+\rangle_{t_1}, \vert 1' \rangle = \vert-\rangle_{t_1}, \vert 2' \rangle = \vert+\rangle_{t_2}, \vert 3' \rangle = \vert-\rangle_{t_2} \}$.

\textbf{Basis (iii)}, using $\{ \vert H \rangle_{c},\vert V \rangle_{c}; \vert + \rangle_{t}, \vert - \rangle_{t}\}$. The input/output probability matrix reads:
\begin{equation}
P_{p}^{(iii)} = \begin{pmatrix}
\frac{15+p}{4(9-5p)} & \frac{3(1-p)}{4(9-5p)} &\frac{9(1-p)}{4(9-5p)} & \frac{9(1-p)}{4(9-5p)} \\
\frac{3(1-p)}{4(9-5p)} &\frac{15+p}{4(9-5p)} &\frac{9(1-p)}{4(9-5p)} & \frac{9(1-p)}{4(9-5p)} \\
\frac{3(1-p)}{4(9-5p)} &\frac{3(1-p)}{4(9-5p)} &\frac{21-5p}{4(9-5p)} &\frac{9(1-p)}{4(9-5p)} \\
\frac{3(1-p)}{4(9-5p)} &\frac{3(1-p)}{4(9-5p)} &\frac{9(1-p)}{4(9-5p)} &\frac{21-5p}{4(9-5p)}
\end{pmatrix}.
\end{equation}
The rows correspond to the following input two-photon states: $\{ \vert +_t H_c\rangle, \vert +_t V_c\rangle, \vert -_t H_c\rangle, \vert -_t V_c\rangle\}$. The columns correspond to the following single-photon output states: $\{ \vert 0'' \rangle = \vert H\rangle_{t_+}, \vert 1'' \rangle = \vert V\rangle_{t_+}, \vert 2'' \rangle = \vert H\rangle_{t_-}, \vert 3'' \rangle = \vert V\rangle_{t_-} \}$, where the output modes $t_+$ and $t_-$ are defined according to the rule $\vert H\rangle_{t_\pm}=(\vert H\rangle_{t_1}\pm \vert H\rangle_{t_2})/\sqrt{2}$ and analogously for the other polarizations.

\textbf{Basis (iv)}, using $\{ \vert + \rangle_{c},\vert - \rangle_{c}; \vert + \rangle_{t}, \vert - \rangle_{t}\}$. The input/output probability matrix reads:
\begin{equation}
P_{p}^{(iv)} = \begin{pmatrix}
\frac{3+p}{6-2p} & 0 & \frac{3(1-p)}{6-2p} & 0 \\
0 & \frac{3+p}{12-8p} & \frac{9(1-p)}{12-8p} & 0 \\
0 & \frac{3(1-p)}{12-8p} & \frac{2(3-p)}{12-8p} & \frac{3(1-p)}{12-8p} \\
0 & 0 &\frac{3(1-p)}{6-2p} & \frac{3+p}{6-2p} 
\end{pmatrix}.
\end{equation}
The rows correspond to the following input two-photon states: $\{ \vert +_t +_c \rangle, \vert +_t -_c \rangle, \vert -_t +_c\rangle, \vert -_t -_c\rangle\}$. The columns correspond to the following single-photon output states: $\{ \vert 0''' \rangle = \vert+\rangle_{t_+}, \vert 1''' \rangle = \vert-\rangle_{t_+}, \vert 2''' \rangle = \vert+\rangle_{t_-}, \vert 3''' \rangle = \vert-\rangle_{t_-} \}$.

%
%
\begin{figure}[ht!]
\centering
\includegraphics[width=0.235\textwidth]{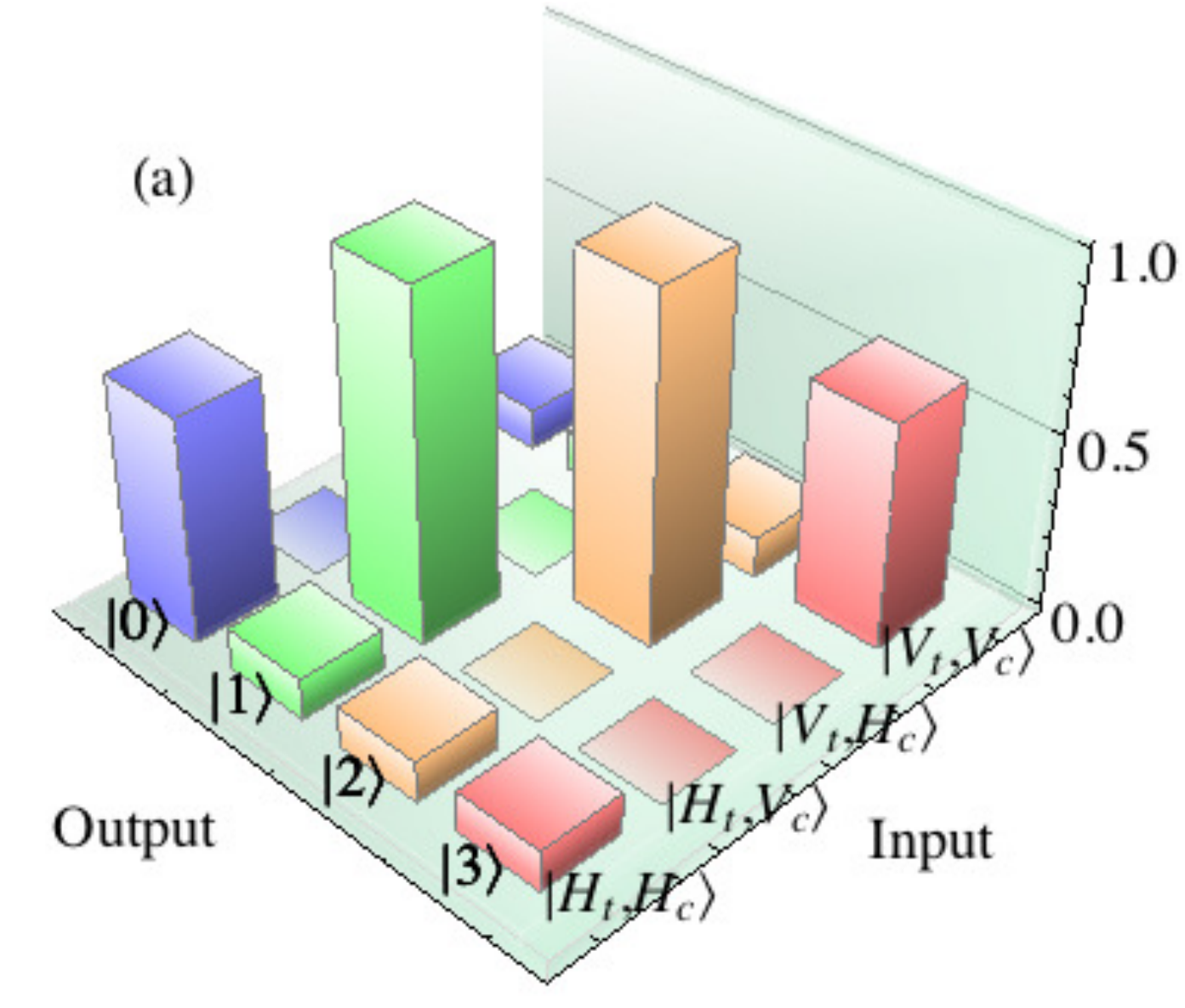}
\includegraphics[width=0.235\textwidth]{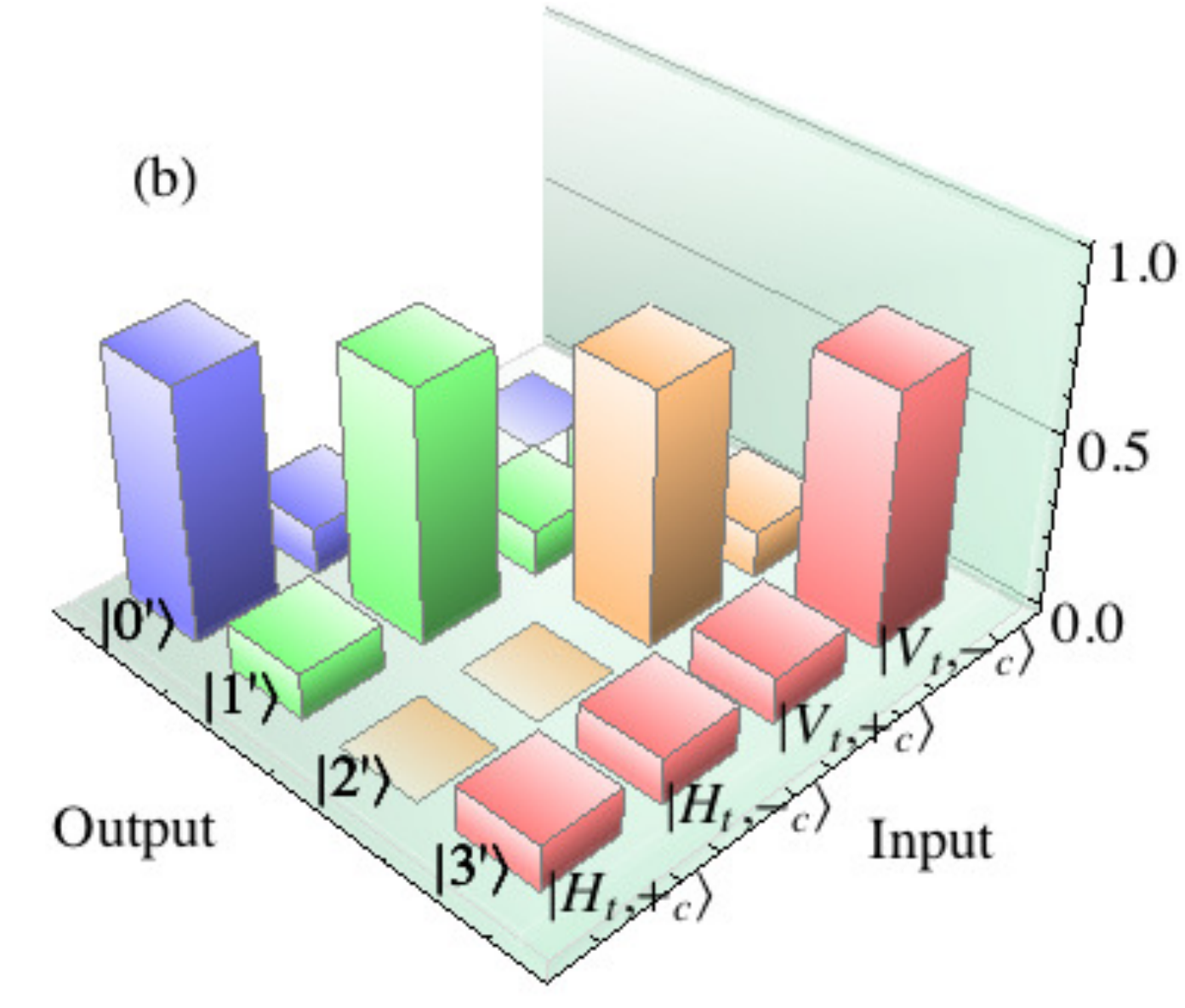}
\includegraphics[width=0.235\textwidth]{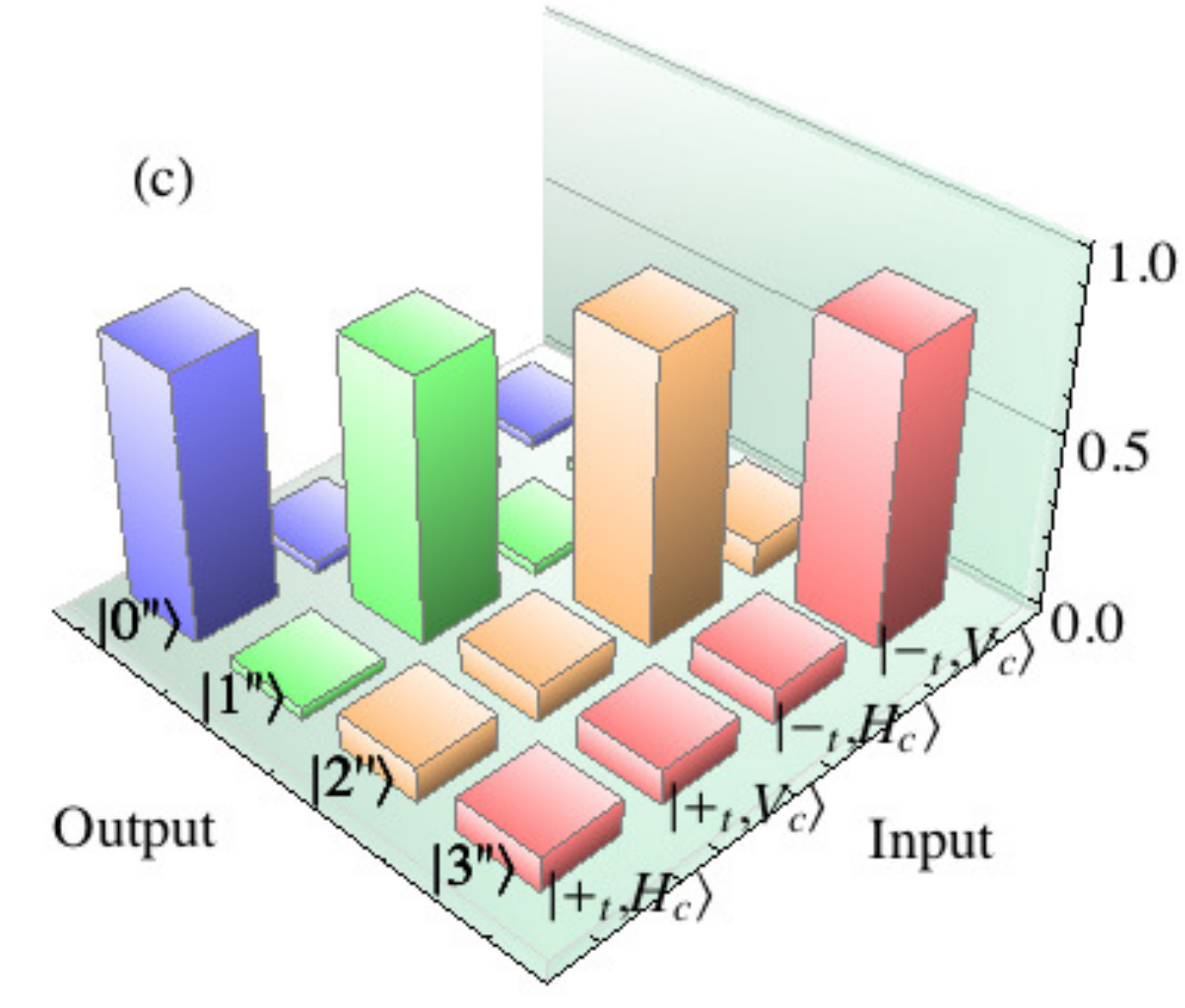}
\includegraphics[width=0.235\textwidth]{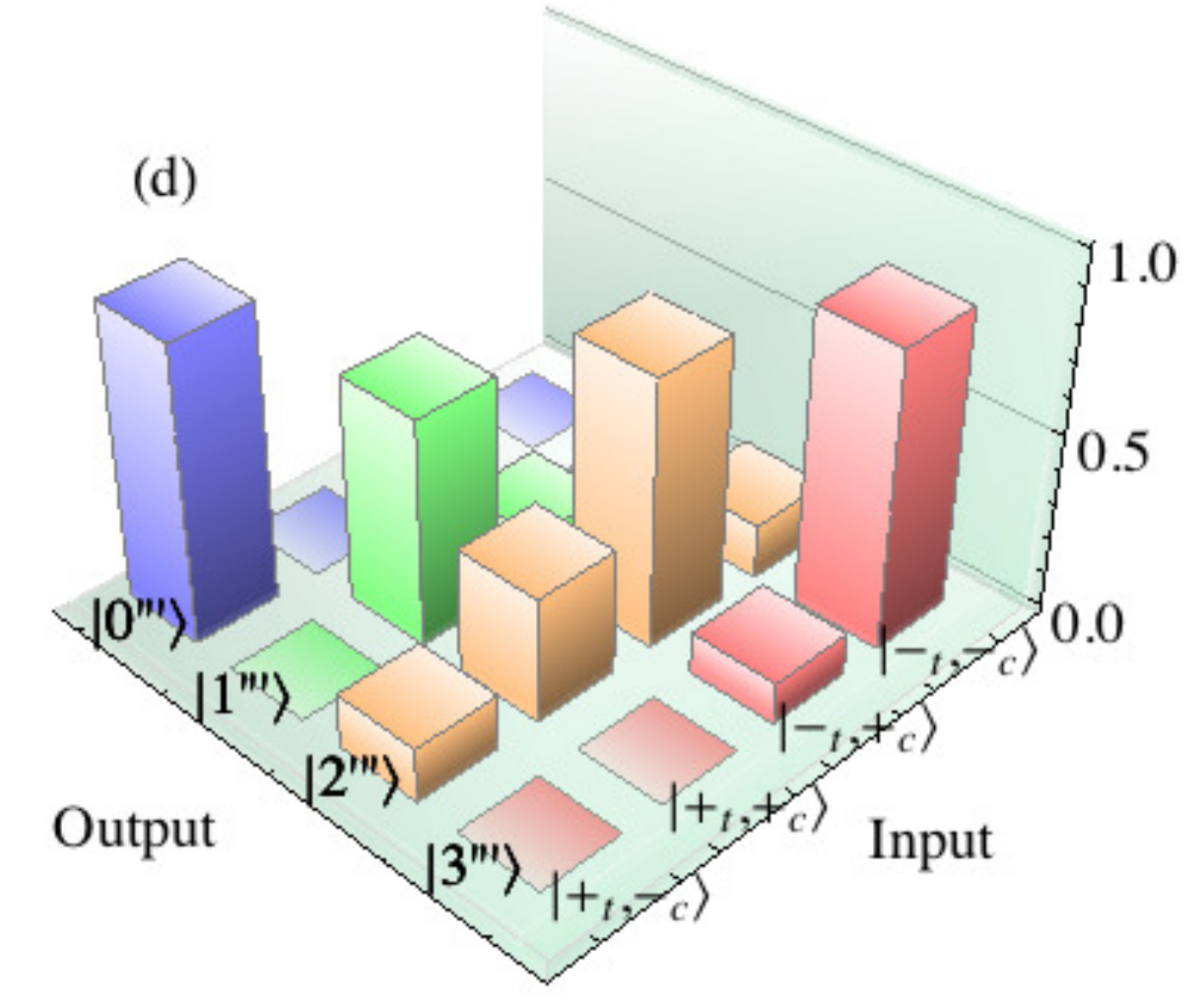}
\caption{Input/output probabilities for the different bases tested in our experiment, for a degree of indistinguishability $p=0.77$ of the photon pairs: (a) basis (i); (b) basis (ii); (c) basis (iii); and (d) basis (iv).}
\label{fig:model}
\end{figure}
%
%
%
\begin{figure}[ht!]
\centering
\includegraphics[width=0.35\textwidth]{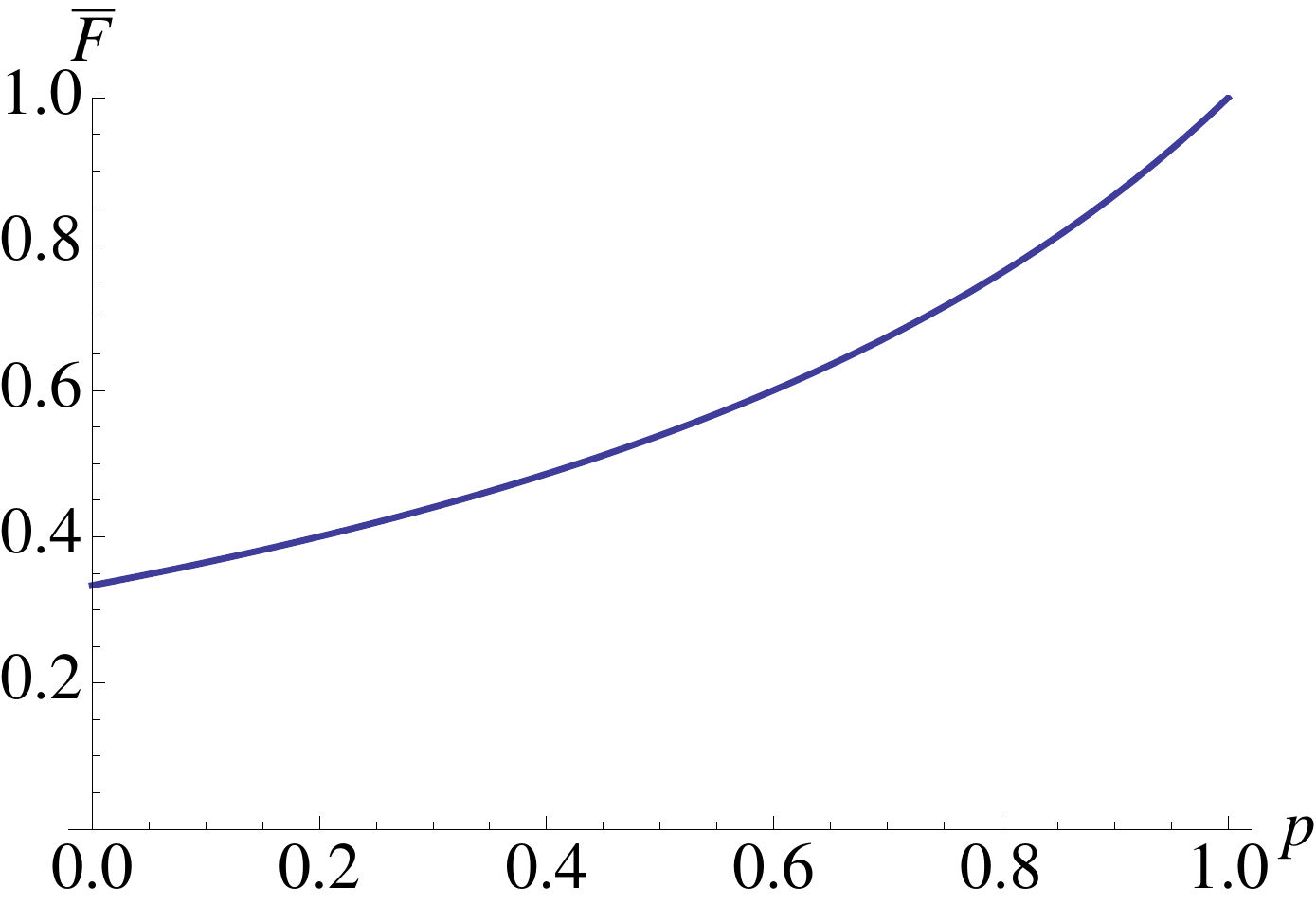}
\caption{Average quantum-state fusion fidelity in the four bases tested in our experiment, as a function of the indistinguishability parameter $p$.}
\label{fig:model_av_fidelity_p}
\end{figure}

The value of the indistinguishability parameter $p$ applicable to our photon source has been obtained by optimizing the agreement between the experimental results and the model predictions in the case of basis (ii), which shows a marked non-uniformity of the fidelity. We obtained $p=0.77$, in good agreement with the results of the direct Hong-Ou-Mandel visibility measurement.

In Fig.\ \ref{fig:model} we report the theoretical input/output probabilities for the states of the bases (i)-(iv) and for a degree of indistibuishability $p=0.77$ of the photon pairs. From these results we can also evaluate the average fidelity of the qubit fusion process for the entire chosen set of 16 input states, which reads
\begin{equation}
\overline{F} = \frac{3+p}{9-5p}.
\end{equation}
and is shown in Fig.\ \ref{fig:model_av_fidelity_p}.

%
%
%
\begin{figure}
\begin{center}
\includegraphics[width=8cm]{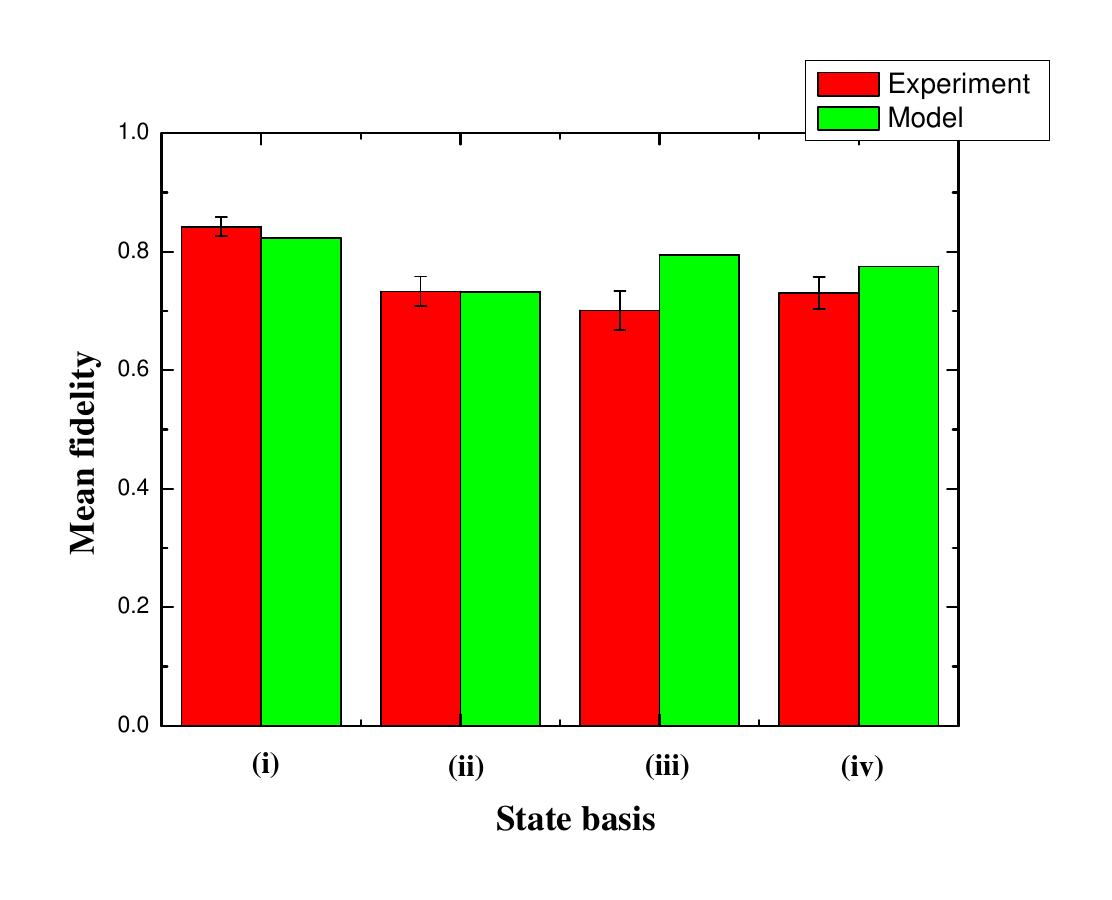}
\end{center}
\caption{Average measured fidelities (red bars) for the four photon-pair bases listed in Fig.\ \protect\ref{figresultsstates} compared with the predicted fidelity according to our model (green bars).} \label{figfidelitymean}
\end{figure}

Finally, in Fig.\ \ref{figfidelitymean} we report the average predicted fidelities for each tested basis compared with the results of our experiment.

\end{document}